\documentclass{sig-alternate-2013}

\newfont{\mycrnotice}{ptmr8t at 7pt}
\newfont{\myconfname}{ptmri8t at 7pt}
\let\crnotice\mycrnotice%
\let\confname\myconfname%

\usepackage[
pass,
]{geometry}

\permission{Permission to make digital or hard copies of part or all of this work for personal or classroom use is granted without fee provided that copies are not made or distributed for profit or commercial advantage and that copies bear this notice and the full citation on the first page. Copyrights for third-party components of this work must be honored. For all other uses, contact the Owner/Author(s). Copyright is held by the owner/author(s).}
\conferenceinfo{SIGIR'15,}{August 09-13, 2015, Santiago, Chile.} 
\copyrightetc{ACM \the\acmcopyr}
\crdata{978-1-4503-3621-5/15/08. \\
	DOI: http://dx.doi.org/10.1145/2766462.2767870 }

\usepackage{epsfig}
\usepackage{comment}

\newcommand{\superscript}[1]{\ensuremath{^{\textrm{#1}}}}

\def\wu{\superscript{*}}
\def\wg{\superscript{\dag}}

\begin{document}

\title{DINFRA: A One Stop Shop for Computing Multilingual Semantic Relatedness}

\numberofauthors{1}
\author{
\alignauthor Siamak Barzegar\wu, Juliano Efson Sales\wu, Andre Freitas\wg, Siegfried Handschuh\wg, \\ Brian Davis\wu\\
\affaddr{{\wu} Insight Centre for Data Analytics, NUI Galway, Ireland}\\
\email{Firstame.Lastname@insight-centre.org}\\
\affaddr{{\wg} Department of Computer Science and Mathematics, University of Passau, Germany}\\
\email{Firstame.Lastname@uni-passau.de}
}

\maketitle
\begin{abstract}
This demonstration presents an infrastructure for computing multilingual semantic relatedness and correlation for twelve natural languages by using three distributional semantic models (DSMs). Our demonsrator - DInfra (Distributional Infrastructure) provides researchers and developers with a highly useful platform for  processing  large-scale corpora and conducting experiments with distributional semantics.
We integrate several multilingual DSMs in our webservice so end user can obtain a result without worrying about the complexities involved in building DSMs. Our webservice allows the users to have easy access to a wide range of comparisons of DSMs with different parameters. In addition, users can configure and access DSM parameters using a easy to use API.

\end{abstract}

.


\section{Introduction}
Dinfra is an implementation of Explicit Semantic Analysis (ESA), Latent Semantic Analysis (LSA) and Random Indexing based on the EasyESA \cite{carvalho2014easyesa}  and S-Space \cite{jurgens2010s} . It runs as a JSON\footnote{JSON - Java Script Object Notation} webservice, which allows users to submit queries for similar terms in a multilingual fashion bases on a semantic relatedness measure which use Spearman's correlation to test relatedness scores. \\
The Dinfra webservice allows the user to obtain semantic similarity using Spearman correlation for \textbf{12} natural languages\footnote{English, Portuguese, German, Spanish, French, Swedish, Italian, Dutch, Chinese, Russian, Arabic and Persian}. Our service can be tested online\footnote{http://vmdgsit04.deri.ie:8008}.  It includes \textbf{two} components: 1- Semantic Relatedness (Figure \ref{fig2}) that calculates the words similarity, 2- Correlation (Figure \ref{fig3}) that calculates the spearman's rank correlation.

\section{Related Work}

Ferret \cite{ferret2010testing} tested corpust-based approaches for measuring semantic similarity. He also chose to use limited means because of deficit of linguistic tools are not, or at least freely available, for all popular languages. Bullinaria et al. \cite{bullinaria2012extracting,bullinaria2007extracting} have built semantic vectors from very small co-occurrence windows, together with a cosine distance measure, stopwords, word stemming, and dimensionality reduction using singular value decomposition to improve performance. The BNC and (British National Corpus)\footnote{http://www.natcorp.ox.ac.uk/} and ukWaC\footnote{2 billion word corpus constructed from the Web limiting the crawl to the .uk domain and using medium-frequency words from the BNC} corpus were used In \cite{bullinaria2007extracting} and \cite{bullinaria2012extracting}, respectively.

\section{System Demonstration}

Three word similarity datasets WordSim353 (W353), the Rubenstein \& Goodenough (RG) (1965) and Miller \& Charles (MC) (1991) have been used in Dinfra. All these datasets consist of human similarity ratings for word pairings.
We also consider Wikipedia\footnote{http://en.wikipedia.org/wiki/Wikipedia:Database\_download} corpus the years (2006, 2008, 2014) and ukWaC \cite{baroni2009wacky} corpus from which to build the vectors.

In Dinfra, three DSMs were instantiated. Latent Semantic Analysis (LSA) \cite{landauer1998introduction}, Random Indexing (RI) \cite{sahlgren2001vector} and Explicit Semantic Analysis (ESA) \cite{gabrilovich2007computing}. The different combinations of DSMs and corpora were evaluated for the computation of semantic similarity and relatedness measures. 

\begin{figure}
	\centering
	\epsfig{file=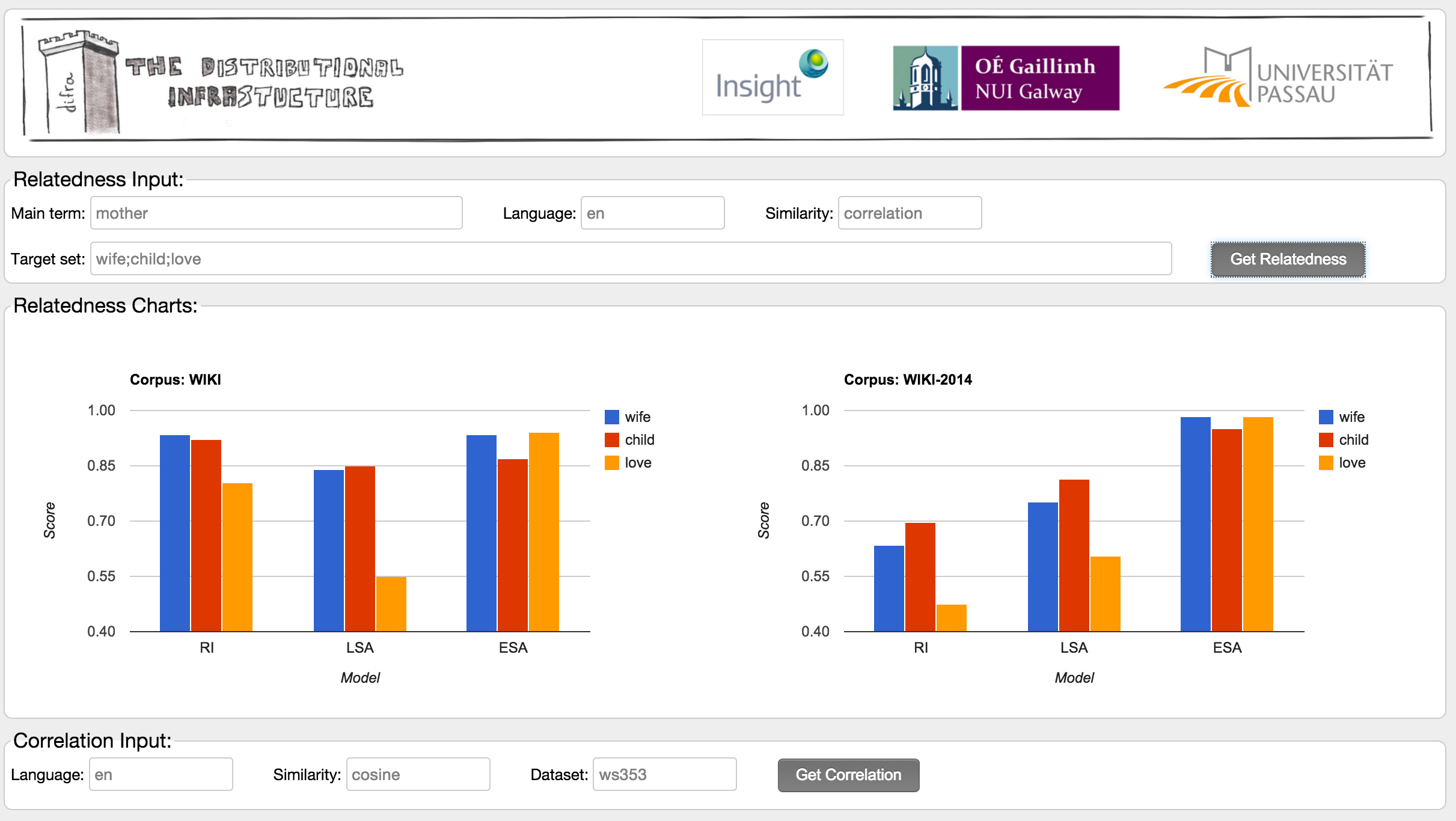 , height=2.3in, width=3.3in}
	\caption{Semantic Relatedness Component}
		 	\label{fig2}
\end{figure}

For  the semantic relatedness component (Figure \ref{fig2}), four \textit{parametrs} such as main term, target set, language and similarity measure are used. The user can compare target words to main word with three similarity measures in \textbf{twelve} different languages.
For the example, we compared (\textit{Wife}, \textit{Child} and \textit{love}) with \textit{mother}, also we used the \textit{Correlation}\footnote{A mean-adjusted version of Cosine as defined in \cite{kiela2014systematic}} measure, Figure \ref{fig2} shows the results that is returned by our webservice.
The semantic relatedness measure is a real number within the [0,1] interval, representing the degree of semantic proximity between two terms. Semantic relatedness can be used for semantic matching in the context of the development of semantic systems such as question answering, text entailment, event matching and semantic search\cite{carvalho2014easyesa} and also for entity/word sense disambiguation tasks.

\begin{figure}[h]
	\centering
	\epsfig{file=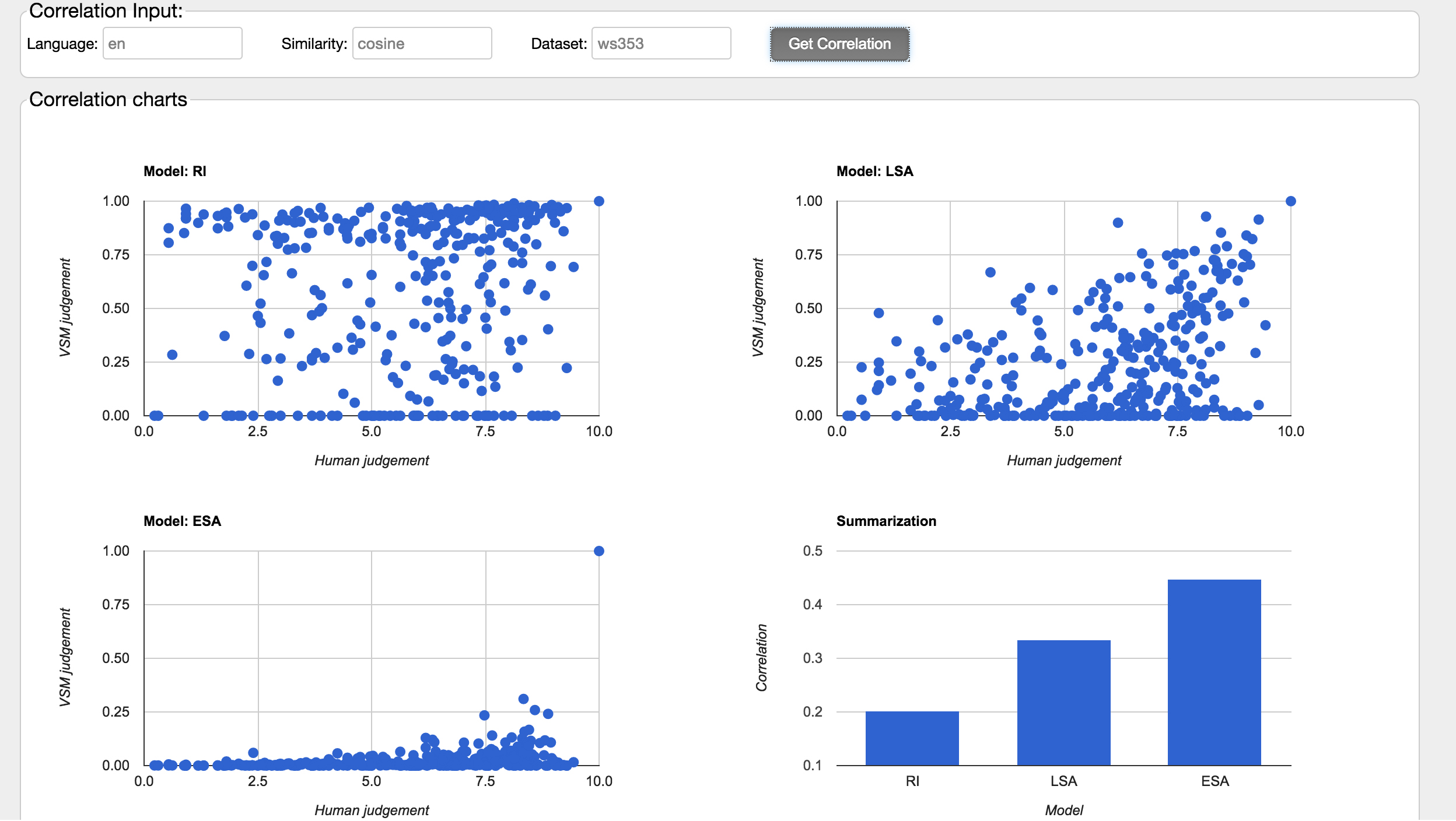 , height=2.3in, width=3.3in}
	\caption{Correlation Component}
	\label{fig3}
\end{figure}

The correlation component (Figure \ref{fig3}) calculates the Spearman's rank correlation for the three similarity datasets, twelve different languages and three similarity measures (Cosine, Euclidean distance, Correlation)\footnote{See \cite{kiela2014systematic} page 3 for definitions of these
similarity measures.}.

All three datasets WS353, RG and MC were translated and localised by native speakers for each of the target 11 languages. More importantly the localised datsets for each language underwent a linguistic
quality assurance by a well know localisation company.  Hence, we are confident that our localised datasets per language are of high translated quality.

\vfill\eject
\section{Acknowledgments}
This publication has emanated from research conducted
with the financial support of Science Foundation Ireland
(SFI) under Grant Number SFI/12/RC/2289.

We would like in particular to thank Alexandros Poulis and Juha Vilhunen from the Lionbridge Natural Language Solutions ensuring the production word of high quality translations for our similarity datasets.
\bibliographystyle{abbrv}
\bibliography{sigproc}

\end{document}